\documentclass[12pt]{article}
\usepackage{pic02}
\usepackage{hyperref}
\usepackage{url}
\usepackage{graphicx}

\begin{document}

\title {\bf GLAST DARK MATTER SEARCH} \author{ Lawrence Wai \\ {\em
Stanford Linear Acclerator Center, Stanford, California} \\
Representing the GLAST (LAT) Collaboration
\footnote{www-glast.stanford.edu} } \maketitle

\baselineskip=14.5pt
\begin{abstract}
The GLAST Large Area Telescope~\cite{GLAST}, scheduled for launch in
2006, is a next generation space based gamma ray telescope which will
improve in point source sensitivity by a factor of 30 over that of
EGRET~\cite{EGRET} below 10~GeV, and extend beyond EGRET up to
300~GeV.  Thus GLAST offers a unique opportunity to discover WIMP dark
matter through precision studies of gamma rays produced in pair
annihilations.  The most dense region of dark matter in our galaxy is
currently thought to occur at the center; in particular, dark matter
should concentrate within 3~pc of the putative supermassive black hole
located at the SgrA* radio source~\cite{GS}.  In fact, the 2nd and 3rd
EGRET catalogs contain a significant point source coincident with the
Milky Way galactic center within a resolution of
12~arcminutes~\cite{MH}.  The EGRET team has determined that the
spectral and temporal characteristics of this point source are
consistent with dark matter WIMP annihilations.  More detailed
analysis \cite{Silk2} has determined that the magnitude and spectrum
of the EGRET source is consistent with relic WIMPs concentrated within
3~pc of the central supermassive black hole.  Furthermore, the SgrA*
radio emission is consistent with the synchrotron radiation expected
from electrons and positrons produced in WIMP annihilations.  If true,
then GLAST should be able to constrain the particle properties of the
postulated WIMP with 1 month of data.
\end{abstract}

\baselineskip=17pt

The accretion of a collisionless collection of non-relativistic
identical particles with asymptotic velocity $v_0$ onto a central star
of mass M was first examined by Zel'dovich and Novikov \cite{ZN}.  In
particular, we can define a critical radius $r_c = {2GM}/{v_0^2}$.
For distances $r>>r_c$, the density varies little from the asymptotic
value; however, for $r<<r_c$ the gravitational potential of the cntral
mass focuses the particles together, thus greatly increasing their
density.  Gondolo and Silk \cite{GS} re-analyzed this situation in the
context of dark matter particles focusing near the putative
supermassive black hole at the center of our galaxy \cite{Ghez}.  For
$M=2.6\times10^6$ solar masses $r_c=3$~pc for an asymptotic particle
dark matter velocity $v_0=90$~km/s.  The density inside of 3~pc is
high enough to produce observable signals from particle dark matter
annihilation.

\begin{figure}[htbp]
\centerline{\hbox{\includegraphics[width=8.0cm]{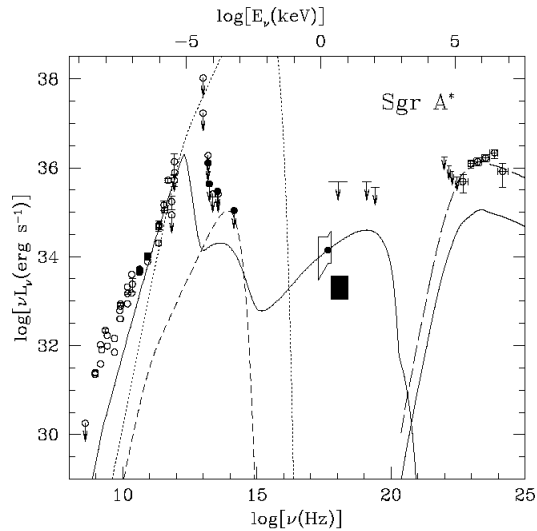}}}
\caption{\it From \cite{Narayan}: circles represent various flux
measurements and upper limits of Sgr A*.  The four peaks from left to
right represent: synchrotron radiation, Compton scattering,
bremsstrahlung, pion production.  The dotted line is the spectrum
corresponding to a standard thin accretion disk with accretion rate
1e-4 solar mass, and the short-dashed line is thin disk with accretion
rate 1e-9 solar mass.  The solid line is the advection dominated
accretion flow (ADAF).  The long dashed line shows ADAF with pion peak
artificially raised by 1 order of magnitude.  Of course, all of these
models apply strictly to non-dark matter accretion scenarios.  The
{\it Chandra} observation in the 2-10~keV band \cite{Baganoff} is
shown as the black box.
\label{sgrastar}}
\end{figure}

\begin{figure}[htbp]
\centerline{\hbox{\includegraphics[width=8.0cm]{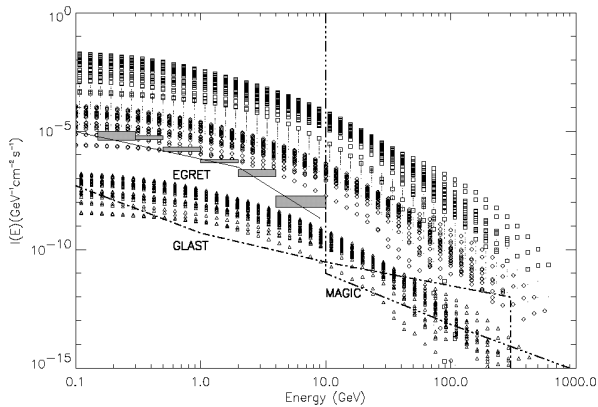}}}
\caption{\it From \cite{Silk2}: EGRET data and expected gamma ray flux
from the galactic center for various SUSY models.  Variation with cusp
slope is also shown: $\gamma=0.05$ (triangles), $\gamma=0.12$
(diamonds), $\gamma=0.2$ (dots), $\gamma=1.0$ (squares) \label{silk2}}
\end{figure}

Multi-wavelength observational data from SgrA* has been compiled by
Narayan {\it et al} for comparison with various (non-dark matter)
accretion models, shown in figure~\ref{sgrastar}.  Silk {\it et al}
\cite{Silk1,Silk2} has re-analyzed this data in terms of
supersymmetric (SUSY) particle dark matter.  A good fit to radio and
gamma ray data has been found for typical particle dark matter
paramaters.  In figure \ref{silk2} we show a comparison of EGRET data,
SUSY model predictions, and GLAST sensitivity.  With one month of
data, GLAST should be able to probe most of the supersymmetric
parameter space.  The final GLAST data sample should obtain an
accuracy of 10~arcseconds, which corresponds to 0.4~pc at the galactic
center.

\end{document}